\documentclass[english,aps,prper,reprint,showpacs,titlepage,longbibliography,nofootinbib,superscriptaddress]{revtex4-2}

\usepackage[T1]{fontenc}	
\usepackage[latin9]{inputenc}	
\usepackage{geometry}		
\usepackage{graphicx}
\usepackage[above,below]{placeins}	
\usepackage{times}
\usepackage{multirow}
\usepackage[dvipsnames]{xcolor}
\usepackage{subcaption}
\usepackage{titlesec}
\usepackage{ifthen}
\usepackage{makecell}
\newenvironment{iquote}
    {\vspace{-.33\baselineskip}\itshape\list{}{\leftmargin=0.15in\rightmargin=0.15in}%
    \item\relax}
    {\endlist\vspace{-.33\baselineskip}}

\newcommand{\ket}[1]{|#1\rangle}

\newcommand{\fracroot}[2]{\ifthenelse{#1=1}{\frac{1}{\sqrt{#2}}}{\sqrt{\frac{#1}{#2}}}}

\usepackage{makecell}

\newcommand{\who}[1]{\textbf{#1\emph{:}}}

\newif\ifshowtimestamp
\showtimestampfalse

\newcommand{\timestamp}[1]{
    \ifshowtimestamp{\textcolor{red}{Timestamp: #1}}\fi}

\usepackage{hyperref}  
\hypersetup{colorlinks=true,urlcolor=blue,citecolor=blue,linkcolor=blue}   
\usepackage{enumitem}        
\setlist{nosep}                 

\usepackage{mathtools}
\usepackage{amssymb}
\usepackage{amsmath}
\usepackage{amsthm}
\usepackage[font=small]{caption}
\usepackage{subcaption}
\usepackage{bbold}
\usepackage{comment}
\usepackage{tabularx}
\usepackage{array}
\usepackage{censor}

\usepackage[dvipsnames]{xcolor}
\usepackage{footmisc}

\usepackage[sort&compress]{natbib}

\begin{document}
\setlist[itemize]{leftmargin=20pt}
\begin{titlepage}

\title{Investigating student interpretations of the differences between classical and quantum computers: Are quantum computers just analog classical computers?}
\author{Josephine C.\ Meyer}
\affiliation{Department of Physics, University of Colorado - Boulder, Boulder, CO 80309}
\author{Gina Passante}
\affiliation{Department of Physics, California State University - Fullerton, Fullerton, CA 92831}
\author{Steven J.\ Pollock}
\author{Bethany R.\ Wilcox}
\affiliation{Department of Physics, University of Colorado - Boulder, Boulder, CO 80309}

\begin{abstract}
Significant attention in the PER community has been paid to student cognition and reasoning processes in undergraduate quantum mechanics. Until recently, however, these same topics have remained largely unexplored in the context of emerging interdisciplinary quantum information science (QIS) courses. We conducted exploratory interviews with 22 students in an upper-division quantum computing course at a large R1 university crosslisted in physics and computer science, as well as 6 graduate students in a similar graduate-level QIS course offered in physics. We classify and analyze students' responses to a pair of questions regarding the fundamental differences between classical and quantum computers. We specifically note two key themes of importance to educators: (1) when reasoning about computational power, students often struggled to distinguish between the relative effects of exponential and linear scaling, resulting in students frequently focusing on distinctions that are arguably better understood as analog-digital than classical-quantum, and (2) introducing the thought experiment of analog classical computers was a powerful tool for helping students develop a more expertlike perspective on the differences between classical and quantum computers.

\clearpage
\end{abstract}

\maketitle
\end{titlepage}

\section{Introduction \& Background}

Fueled by the rapid growth in demand for jobs in the quantum industry as well as funding from the National Quantum Initiative Act of 2018, interdisciplinary quantum information science (QIS) coursework has begun flourishing at all levels across U.S.\ universities \cite{Plunkett:2020, Aiello:2021, Cervantes:2021, Asfaw:2022, Meyer:2022PhysRev}. Such coursework can cover a variety of related topics including quantum computing, quantum sensing, and quantum cryptography. Due to their novelty, little work has been done on such courses by the physics education research (PER) community and other discipline-based education research (DBER) communities that QIS coursework touches. Recognizing the opportunity to institute research-based instructional practices from the beginning, 32 scientists and professionals in QIS and adjacent fields signed an open letter calling for 
the early involvement of education experts in curriculum development \cite{Aiello:2021}.

Much DBER work on QIS education has specifically focused on industry needs for workforce development \cite{Aiello:2021,Fox:2020,Singh:2021,Hughes:2022}, trends in existing coursework \cite{Plunkett:2020,Cervantes:2021,Meyer:2022PhysRev}, or teaching quantum computing to computer science students \cite{Mermin:2003,Grau:2004,Salehi:2021,Seegerer:2021}. Chandralekha Singh's group has pioneered a series of Quantum Interactive Learning Tutorials (QUILTs) to help students in existing quantum mechanics classes learn quantum information processing \cite{Singh:2006,DeVore:2014,DeVore:2020}, and recent work has developed curricular materials for secondary school audiences \cite{Satanassi:2022}.

However, largely missing from the literature so far has been explicit study of student reasoning in the context of QIS courses. Much of what we do know about student learning in QIS contexts comes from studies in traditional upper-division quantum mechanics courses \cite{Singh:2009,Zhu:2012,Kohnle:2015,Passante:2015,Singh:2015,Wan:2019,Li:2021}, though the language and toolkit (e.g.\ quantum gates) of a typical QIS course are different enough that we expect student difficulties and reasoning processes to have only partial overlap. 

Recognizing this gap, we are conducting an ongoing study of student reasoning in upper-division and graduate quantum information and computing courses at a large R1 research university. 
In previous work, we identified student strategies for interpreting quantum states in an upper-division quantum computing course \cite{Meyer:2021}.
This second paper presents preliminary findings on a distinct topic: student interpretations of the difference between quantum and classical computers. 


\section{Significance}

Though the proliferation of QIS courses undoubtedly means that more students will gain exposure to quantum mechanics, it is unclear whether this exposure translates to a deeper appreciation of the physical significance of quantum mechanics. It is entirely conceivable that students could emerge from a quantum computing class with an awareness of \textit{what} a quantum computer can do but with little appreciation of \textit{why} it can solve certain problems more efficiently. For those of us in the PER community, it is valuable to understand what physical implications students are actually taking away from QIS classes and how their interpretations of quantum mechanics compare with experts and with students in traditional quantum mechanics courses. For computer scientists in QIS education, whether students can accurately qualitatively distinguish between classical and quantum computing paradigms is 
of similarly high importance. 

Complicating matters, it may not be realistic to expect non-physics majors in a quantum computing course to solve the same problems or be exposed to the same physical systems as students in a traditional quantum mechanics course. It is thus particularly fruitful to focus on students' qualitative interpretations of the differences between classical and quantum computing paradigms. We argue that students' qualitative interpretations of how quantum and classical computers differ are less susceptible to issues such as notation conventions while closely reflecting the core physics content. 

\begin{table*}[htb!]
    \centering
    \begin{tabular}{c c c c c c}
        \hline
        \hline
        \thead{Semester} & \thead{Level} & \thead{Course Mode} & \thead{Listed Department(s)} & \thead{Interviewees} & \thead{Participation Rate}\\
        \Xhline{2\arrayrulewidth}
        \thead{Spring 2021} & BFY Undergraduate & Remote & Physics, Computer Science & 13 & 19\% \vspace{-2pt}\\ 
        \vspace{-2pt}
        \thead{Fall 2021} & Graduate & In Person & Physics & 6 & 17\%\\ 
        \thead{Spring 2022} & BFY Undergraduate & In Person* & Physics, Computer Science & 9 & 13\%\\ 
        \hline
        \hline
    \end{tabular}
    \vspace{-2pt}
    \caption{Profile courses in which interviewees were recruited. BFY denotes beyond-first-year. All course sections were taught by physicists, except the Spring 2021 course which was co-taught by a computer scientist. Participation rate measures how many students opted to participate in the interview study as a percentage of students who completed the course (defined as completing the final exam in the undergraduate class, or all four homework assignments in the graduate class) and excludes two students in Spring 2021 who did not complete this portion of the interview due to time constraints. *Remote teaching for first two weeks of semester.}
    \label{tab:courses}
\end{table*}

\section{Methodology}

We conducted one-hour Zoom interviews of undergraduate and graduate students from three quantum information courses. The majority of the interview protocol was conducted in a think-aloud format (see e.g.\ Ref.~\cite{Meyer:2021}). The final portion of the interview protocol -- which includes the questions discussed in this paper -- followed a more open-ended approach intended to explore student reasoning. Details 
are summarized in Table~\ref{tab:courses}. Interviews were conducted no earlier than 3 weeks prior to completion of the course. Students were awarded a \$20 gift card for their participation. 

For the subset of the interview protocol analyzed in this paper, students were asked to respond to the following prompts. Interviewees were informed that there was not necessarily a correct answer but that we were interested in their thoughts:

\begin{itemize}
    \item a.\ If you had to pick one thing, what would you say is the key difference between a classical computer and a quantum computer?
    \item b.\ The Summit supercomputer (one of the most powerful supercomputers in the world) possesses 250 petabytes of memory. However, even that amount of memory is only sufficient to store the state of \textasciitilde56 qubits (let alone manipulate them)\footnote{For reference, this estimate of 56 qubits assumes each complex coefficient is represented by two 16-bit floating point values with negligible overhead. Given that the size of the Hilbert space scales exponentially with the number of qubits, this estimate is robust to most reasonable assumptions.}! What is it about a quantum computer that you think makes it so mind-bogglingly difficult to simulate on a classical computer?
\end{itemize}
In spring 2021, we passively collected student responses without much feedback to establish a baseline for student responses. In fall 2021 and spring 2022, the interviewer took a more active role in facilitating the interview, intentionally challenging students' initial conceptions with individually-tailored follow-up questions to elicit cognitive conflict. 


Responses to these questions were initially coded via an \textit{a priori} schema based on the limited number of reasonable answers we expected for this question (superposition, entanglement, some sort of non-quantum property, or miscellaneous), with allowance for the inevitability of novel student responses that could not easily be sorted according to this schema.
In practice, this schema proved too simplistic and 
failed to capture the full implications of students' answers. 
For instance, though both ``entanglement'' and ``superposition'' are arguably expertlike\footnote{We recognize that in a field as diverse as QIS there is not likely to be a single expertlike understanding of quantum computing. The inherently subjective judgment as to whether a given response was more or less expertlike was made by a researcher who has completed a graduate-level QIS course and studies QIS education; it should not be understood as expert consensus.} responses -- indeed, one student, Rowan, called the question ``rude'' for being worded as ``if you had to pick one thing...'' when they wanted to select both! -- we found that some students citing the term ``superposition'' were using it simply to refer to the fact that the quantum state can vary continuously, an explanation that captures only part of the richness of the Bloch sphere. 

Because the initial schema proved unproductive and indicator phrases such as ``superposition'' were found not to be used consistently, we transitioned to a more open-ended analysis to identify emergent themes. In particular, where identifiable, we coded for (1) whether students' explanations for the fundamental differences between classical and quantum computers focused on apparently exponentially-scaling (e.g.\ growth of the Hilbert space with the number of qubits) or polynomially-scaling (e.g.\ the continuous and/or complex nature of the state vector) phenomena, and (2) any marked shifts in students' apparent mental model of the fundamental differences between quantum and classical computation over the course of the interview.

\section{Results and Analysis}

In this section, we discuss two important themes emergent from our analysis of students' discourse regarding the fundamental differences between classical and quantum computers: students' tendency not to distinguish between linear and exponential effects as sources of ``quantum advantage,'' leading to the treatment of the classical-quantum distinction as akin to analog-digital, and the value of the analog classical computer thought experiment in helping students clarify what properties of a quantum computer are truly ``quantum.''


\subsection{Student difficulties with linear vs.\ exponential scaling}
\label{subsec:difficulties}

When asked to identify the fundamental differences between classical and quantum computers, or why simulating a quantum computer on a classical computer requires so much memory, a frequent initial response
was that storing the state of the quantum computer requires additional memory (or is entirely impossible) because the quantum state can only be defined in terms of continuous coefficients vs.\ the discrete 0s and 1s of digital classical computers. Even among students who did mention properties such as entanglement driving exponential growth, the continuous nature of the quantum state vector was frequently treated on equal footing. From an expertlike perspective, these effects are of wholly different scale: representing the continuous nature of the state vector to arbitrary precision on a classical computing architecture merely requires the use of floating-point arithmetic,\footnote{Floating-point arithmetic encodes real numbers in terms of two binary bit strings representing, respectively, the power of 2 and its prefactor. Floating-point arithmetic is widely used in classical computing for continuous quantities, and its precision is limited only by round-off error that can normally be suppressed by storing a sufficient number of bits.} whose memory consumption represents at worst a constant factor as compared to integer-valued data. The number of components needed to represent the state vector -- the dimension of the Hilbert space -- however, scales exponentially with the number of qubits ($N$) due to superposition and entanglement.

It is this exponential scaling with $N$ associated with the dimension of the Hilbert space that makes a quantum computer unique from a classical computer; the number of bits needed to store $N$ floating point numbers scales only as $N$. In computer science, a linear scaling is considered tractable: just double the size of your supercomputer, taking advantage of Moore's law,\footnote{Moore's law is the empirical observation that the number of transistors per volume on a computer chip doubles approximately every 1.5-2 years.} and you can double the size of the input you can process. An exponential scaling is by contrast generally considered intractable, because doubling the size of your supercomputer will have negligible effect on the scale of the input that can be processed in reasonable time.

In this study, we observed that students often struggled to distinguish the magnitude of the effects of linear and exponential scaling. This tendency was especially pronounced among undergraduate students without extensive CS exposure, who were less likely to have been exposed to the concepts of ``big $\mathcal{O}$'' notation\footnote{Big $\mathcal{O}$ notation is a framework used in computer science to analyze the time-complexity of algorithms.} and complexity classes in prior CS coursework. The following representative response to part (a) illustrates this difficulty we observed among students:

\begin{iquote}
    \who{Diana (3rd year UG, appl.\ math)} If a regular computer is digital, then ... a quantum computer is like analog ... a range of values instead of only 0s and 1s, which gives you, like, a lot more ability to have information stored.
    \timestamp{16:24 verified}
\end{iquote}

Diana continues to focus on the linear scaling associated with the analog-digital distinction even as the interviewer calls her attention back to a prior exchange in which floating-point arithmetic
was discussed as a solution for storing continuous values on digital machines:

\begin{iquote}
    \who{Diana} A classical bit ... gives you two options, but a qubit gives you infinitely many options ... \who{Interviewer} So I would agree with that. But again as we've said, like you can use floating point arithmetic to arbitrary precision.
    \who{Diana} ... but for that floating point arithmetic
    ... to [be] like really high precision, you're already using a bunch of bits to do that ...
    \who{Interviewer} Yeah, well, even if we supposedly like dedicated 1000 bits [per floating point value] ... that still is only like 56,000 bits, that's not 250,000 terabytes.
    \who{Diana} Yeah, um \textit{[pauses]} I don't know.
    \timestamp{19:34 verified}
\end{iquote}
In fact, it was not until the interviewer explicitly prompted the term ``entanglement'' and walked Diana through a computation of the dimension of the Hilbert space for 1, 2, 3, and then N qubits that Diana connected quantum advantage to exponential growth of the Hilbert space due to entanglement.

Diana's framing of the difference between a quantum computer and a classical computer as primarily an analog vs.\ digital one indeed suggests a potential interpretation in terms of a resources framework \cite{Hammer:2000} -- analogous to our prior findings regarding student interpretation of quantum states in Ref.~\cite{Meyer:2021} --  though more work is needed to substantiate this claim. It seems plausible that Diana continued to persist in her original explanation because the resource of entanglement had not yet been activated and she was thinking solely of properties of individual qubits.

It is worth emphasizing that even among students who recognized the exponentially growing Hilbert space, this phenomenon was often treated on equal footing with linear effects. Here, Felix immediately recognizes the exponential scaling but then describes how the linear effects make it ``even worse,'' a description that though strictly true seems to overlook the relative magnitudes of the effects:

\begin{iquote}
    \who{Felix (3rd year UG, CS)} It seems to me ... like it's a fundamentally exponential problem ... If you have a 2 qubit system ... your vector describing that system is going to be 4 bits.\footnote{Here it appears Felix is using the terms ``bits'' and ``digits'' when he means vector coefficients.} [For] 3 qubits ... now it's 8 digits long, and so on. And even worse, those digits aren't necessarily binary: 
    ... they can have imaginary components and whatnot!
    \timestamp{24:30 verified}
\end{iquote}

While graduate students often also started out by discussing the continuous/complex nature of the quantum state, the primary difference was that graduate students were typically faster to recognize the dominance of exponential growth when confronted with explicit scaling arguments, suggesting they may have gained greater familiarity with this type of argument from experience:

\begin{iquote}
    \who{Abraham (grad, physics)} [The key difference] is that quantum computers are continuous ... representing a whole bunch of qubits takes like, sets of complex numbers ... [vs] sets of 1s and 0s. \who{Interviewer} We can also represent floating point numbers to arbitrary precision on a classical computer just fine. \who{Abraham} Yeah, but floating point numbers won't grow like exponentially!
    \timestamp{44:59 verified}
\end{iquote}
At this point, Abraham immediately jumped into talking about the scaling of entangled vs.\ non-entangled states, demonstrating an ability to shift away from an explanation that failed to scale satisfactorily and to seek out and activate new resources as necessary.


\subsection{The analog classical computer as a thought experiment}

As discussed in Sec.~\ref{subsec:difficulties}, a common tendency among students was to attribute the fundamental difference between classical and quantum computers to the continuous nature of the state vector. This answer is appealing because it is indeed one of the most obvious consequences of quantum superposition: the state of a classical computer is perfectly represented by a discrete bit-string, while the state of a quantum computer must be specified in terms of continuous (complex) variables. However, as discussed, classical computers can efficiently simulate a continuous system to arbitrary precision using floating-point arithmetic. In other words, there is nothing fundamentally quantum about a computer whose state is continuous; focusing on the continuous state of a quantum computer is, by itself, arguably better understood as an analog-digital than as a quantum-classical distinction.

During one of the Fall 2021 graduate student interviews, the interviewer spontaneously posed a thought experiment while discussing this interview problem with a student: what if we had an analog classical computer, that instead of representing bits digitally as 0 or 1 uses ``analog bits'' whose voltage can continuously vary between 0 V and 1 V? How, if at all, would a quantum computer be different from this analog classical computer\footnote{As it turns out, many early classical computers actually \textit{were} designed around analog logic before digital logic won out technologically \cite{Gregersen:2021}.}? This thought experiment was 
repeated in some undergraduate interviews in Spring 2022 in response to the following feedback from interviewee Zach and once again prompted several students to develop more expertlike explanations for the differences between classical and quantum computers:

\begin{iquote}
\who{Zach (4th year UG, physics)} I wish I had a better way to kind of visualize what was physically going on in quantum computing vs.\ in classical. 
\timestamp{37:08 verified}
\end{iquote}

The following exchange illustrates the effectiveness of this thought experiment in helping students develop more expertlike responses by distinguishing between truly quantum phenomena (e.g.\ entanglement) and analog/digital distinctions:

\begin{iquote}
    \who{Brandon (2nd year UG, physics/astronomy)} Instead of just ... your two states, now you have more potential values ... I think of quantum computing as that we can have the sort of in-between bits ... \who{Interviewer} Yeah, ... is that any different than if I had an analog classical computer, that rather than having just 0 and 1 could like have any range of voltages in between? Because like those do exist, they're just not common anymore.
    \who{Brandon} I didn't actually know about that! I feel like that it's still different, because ... [references prior interview problem about an entangled state] we can have ... entanglement between states as well ... as far as I know that would still be a unique point with the quantum computers, that we not only have superpositions but we can have measurements on one qubit affecting another.
    \timestamp{17:13 verified}
\end{iquote}

Here, Brandon's initial response mirrored most other students' in focusing on the continuous nature of quantum states in a mostly analog-digital sense (what he calls ``more potential values''). However, as soon as the interviewer brought up the existence of an analog classical computer, Brandon almost immediately made the connection back to a prior problem he had solved regarding entanglement and identified entanglement as an essential quantum property that could not be understood through the lens of the analog classical computer.


\section{Conclusions}

The analog classical computer as a point of contrast for understanding the difference between classical and quantum computers is valuable precisely because it provides a sort of intermediate frame of reference for comparison. It is intuitive that an analog classical computer is not necessarily any more powerful than a digital computer, since analog vs.\ digital classical data is a familiar distinction in common real-world applications like audio recording. Yet quantum and analog classical computers are similar in a striking way: both store and manipulate data in continuous rather than discrete states, such that an analog classical computer can appear to reproduce some aspects of quantum superposition while failing to capture either entanglement or the full richness of the Bloch sphere. We believe the model of the analog classical computer -- coupled with an explicit discussion of floating point arithmetic as a way to efficiently represent continuous quantities digitally to high precision -- will be a valuable tool moving forward for quantum computing educators.

Another useful finding for educators and discipline-based education researchers alike is that students may 
conceive of states that experts understand as superpositions of $\ket{0}$ and $\ket{1}$ in a sense more akin to an analog-digital rather than classical-quantum distinction. States like $\ket{\pm}$ and $\ket{\pm i}$ are not ``in between'' $\ket{0}$ and $\ket{1}$ in the same sense that 0.5 V is between 0 V and 1 V in an analog classical device; the truly quantum phenomenon of superposition cannot fully be reduced to an analog-digital distinction. Just because a student uses a term such as ``superposition'' or ``entanglement'' to describe how a quantum computer differs from a classical computer does not imply that the student has developed an equally expertlike understanding of the distinction between classical and quantum computing paradigms.

This study suggests several important avenues for future research in the PER and adjacent DBER communities. Curriculum development efforts contrasting the effects of linear (polynomial) and exponential growth would have value beyond computation, since conceptualizing the difference in scale between polynomial and exponential growth is a necessary skill for both computer scientists and physicists alike. The analog classical computer thought experiment -- and distinguishing the analog classical computer from a quantum computer -- may merit developing a dedicated research-based instructional instrument such as a tutorial. More work is also needed to probe what students actually mean when they use terms like ``superposition'' and ``entanglement''; at this juncture all we can observe is that students' invocation of these quantum-specific terms does not imply a mental model of quantum computation commensurate with how experts understand these terms. Further work is needed to explore students' mental models of quantum computation in greater depth, with a focus on how these models are formed and how misleading conceptions arising from them can be challenged. 


\section{Acknowledgments}
Special thanks to Giaco Corsiglia for assistance with interviews, Monika Schleier-Smith and Victoria Borish for inspiring part (b) of the interview question, and our student interviewees. This work is funded by the CU Dept.\ of Physics, the NSF GRFP, and NSF Grants No.'s 2012147 and 2011958.



\clearpage
\begin{thebibliography}{99}

\bibitem{Plunkett:2020}
T.\ Plunkett, T.\ Frantz, H.\ Khatri, P.\ Rajendran, and S.\ Midha. A survey of educational efforts to accelerate a growing quantum workforce. In \textit{Proceedings of the 2020 IEEE International Conference on Quantum Computing and Engineering}, 330-336 (2020).

\bibitem{Aiello:2021}
C.\ Aiello \textit{et al.} Achieving a quantum smart workforce. \textit{Quantum Sci. Technol.} \textbf{6}, 030501 (2021).

\bibitem{Cervantes:2021}
B.\ Cervantes, G.\ Passante, B.\ Wilcox, and S.\ Pollock. An overview of quantum information science courses at US institutions. In \textit{Proceedings of the 2021 Physics Education Research Conference}, 93-98 (2021).

\bibitem{Asfaw:2022}
A.\ Asfaw \textit{et al.} Building a quantum engineering undergraduate program. \textit{IEEE Trans. Ed.} \textbf{65}(2), 220-242 (2022).

\bibitem{Meyer:2022PhysRev}
J.\ Meyer, G.\ Passante, S.\ Pollock, and B.\ Wilcox. Today's interdisciplinary quantum information classroom: Themes from a survey of quantum information science instructors. \textit{Phys Rev Phys Educ Res} \textbf{18}, 010150 (2022).

\bibitem{Fox:2020}
M.\ Fox, B.\ Zwickl, and H.\ Lewandowski. Preparing for the quantum revolution: What is the role of higher education? \textit{Phys Rev Phys Educ Res} \textbf{16}, 020131 (2020).

\bibitem{Singh:2021}
C.\ Singh, A.\ Asfaw, and J.\ Levy. Preparing students to be leaders of the quantum information revolution. \textit{Physics Today} (2021).

\bibitem{Hughes:2022}
C.\ Hughes, D.\ Finke, D.\ German, C.\ Merzbacher, P.\ Vora, and H.\ Lewandowski. Assessing the needs of the quantum industry. \textit{IEEE Trans. Ed.} \textbf{65}(2), 220-242 (2022).

\bibitem{Mermin:2003}
N.\ Mermin. From Cbits to Qbits: Teaching computer scientists quantum mechanics. \textit{Am J Phys} \textbf{71}, 23-30 (2003).

\bibitem{Grau:2004}
B.\ Grau. How to teach basic quantum mechanics to computer scientists and electrical engineers. \textit{IEEE Trans. Ed.} \textbf{47}(2), 220-226 (2004).

\bibitem{Salehi:2021}
\:{O}.\ Salehi, Z.\ Seskir, and \.{I}.\ Tepe. A computer science-oriented approach to introduce quantum computing to a new audience. \textit{IEEE Trans. Ed.} \textbf{65}(1), 1-8 (2022).

\bibitem{Seegerer:2021}
S.\ Seegerer, T.\ Michaeli, and R.\ Romeike. Quantum computing as a topic in computer science education. In \textit{Proceedings of the 16th Workshop in Primary and Secondary Computing Education} (2021).

\bibitem{Singh:2006}
C.\ Singh. Helping students learn quantum mechanics for quantum computing. In \textit{Proceedings of the 2006 Physics Education Research Conference}, 42-45 (2007).

\bibitem{DeVore:2014}
S.\ DeVore and C.\ Singh. Development of an interactive tutorial on quantum key distribution. In \textit{Proceedings of the 2014 Physics Education Research Conference} (2015).

\bibitem{DeVore:2020}
S.\ DeVore and C.\ Singh. Interactive learning tutorial on quantum key distribution. \textit{Phys Rev Phys Educ Res} \textbf{16}, 010126 (2020).

\bibitem{Satanassi:2022}
S.\ Satanassi, E.\ Ercolessi, and O.\ Levrini. Designing and implementing materials on quantum computing for secondary school students: The case of teleportation. \textit{Phys Rev Phys Educ Res} \textbf{18}, 010122 (2022).

\bibitem{Singh:2009}
C.\ Singh and G.\ Zhu. Cognitive issues in learning advanced physics: An example from quantum mechanics. In 
\textit{Proceedings of the 2009 Physics Education Research Conference}, 63-66 (2009).

\bibitem{Zhu:2012}
G.\ Zhu and C.\ Singh. Surveying students' understanding of quantum mechanics in one spatial dimension. \textit{Am J Phys} \textbf{80}, 252-259 (2012).

\bibitem{Kohnle:2015}
A.\ Kohnle and E.\ Deffebach. Investigating student understanding of quantum entanglement. In \textit{Proceedings of the 2015 Physics Education Research Conference}, 171-174 (2015).

\bibitem{Passante:2015}
G.\ Passante, P.\ Emigh, and P.\ Shaffer. Student ability to differentiate between superposition states and mixed states in quantum mechanics. \textit{Phys Rev ST Phys Educ Res} \textbf{11}, 020135 (2015).

\bibitem{Singh:2015}
C.\ Singh and E.\ Marshman. Review of student difficulties in upper-level quantum mechanics. \textit{Phys Rev ST Phys Educ Res} \textbf{11}, 020117 (2015).

\bibitem{Wan:2019}
T. Wan, P. Emigh, and P. Shaffer. Probing student reasoning in relating relative phase and quantum phenomena. \textit{Phys Rev Phys Educ Res} \textbf{15}, 020139 (2019).

\bibitem{Li:2021}
Y. Li, A. Kohnle, and G. Passante. Student difficulties with quantum uncertainty in the context of discrete probability distributions. In \textit{Proceedings of the 2021 Physics Education Research Conference}, 227-232 (2021).

\bibitem{Meyer:2021}
J.\ Meyer, G.\ Passante, S.\ Pollock, M.\ Vignal, and B.\ Wilcox. Investigating students' strategies for interpreting quantum states in an upper-division quantum computing class. In \textit{Proceedings of the 2021 Physics Education Research Conference}, 289-294 (2021).





\bibitem{Hammer:2000}
D. Hammer, Student resources for learning introductory physics, \textit{Am J Phys} \textbf{68}, (2000).

\bibitem{Gregersen:2021}
E.\ Gregersen. Analog computer. \textit{Encyclopaedia Britannica} (2021).



\end{thebibliography}
\end{document}